\documentclass[structabstract]{aa}  
\usepackage{natbib}
\bibpunct{(}{)}{;}{a}{}{,} 
\usepackage{graphicx}
\usepackage{txfonts}
\usepackage{lscape}
\newcommand{\suzaku}{{\it Suzaku}}
\newcommand{\xmm}{{\it XMM-Newton}}

\newcommand{\chandra}{{\it Chandra}}

\newcommand{\nh}{\mbox {$N_{\rm H}$}}
\newcommand{\hi}{H\,{\sc i}}
\newcommand{\hii}{H\,{\sc ii}}

\newcommand{\about}{$\sim$\kern.03em}
\def\la{\mathrel{\hbox{\rlap{\hbox{\lower4pt\hbox{$\sim$}}}\hbox{$<$}}}}

\begin{document}

   \title{\xmm\ observation of the Galactic supernova remnant W51C
(G49.1--0.1)
          \thanks{Based on observations obtained with \xmm, an ESA 
                  science mission with instruments and contributions 
                  directly funded by ESA Member States and NASA.}
         }

   \subtitle{}

   \titlerunning{\xmm\ observation of the SNR W51C}

   \author{
           Manami Sasaki
          \and
           Cornelia Heinitz
          \and
           Gabriele Warth
          \and
           Gerd P\"uhlhofer
          }

   \institute{Institut f\"ur Astronomie und Astrophysik, 
              Universit\"at T\"ubingen,
              Sand 1, 
              D-72076 T\"ubingen, Germany,
              \email{sasaki@astro.uni-tuebingen.de}
             }

  \date{Received Nov.\ 28, 2013; accepted Jan.\ 15, 2014}

 
  \abstract
   {The supernova remnant (SNR) \object{W51C} is a Galactic object located
in a strongly inhomogeneous interstellar medium with signs of an
interaction of the SNR blast wave with dense molecular gas.}
   {Diffuse X-ray emission from the interior of the SNR can reveal
element abundances in the different emission regions and shed light
on the type of supernova (SN) explosion and its progenitor.
The hard X-ray emission helps to identify possible candidates for a 
pulsar formed in the SN explosion and for its pulsar wind nebula (PWN).}
   {We have analysed X-ray data obtained with \xmm. Spectral analyses
in selected regions were performed.}
   {Ejecta emission in the bright western part of the SNR, located
next to a complex of dense molecular gas, was confirmed. The 
Ne and Mg abundances suggest a massive progenitor with a mass
of $> 20 M_{\sun}$. Two extended
regions emitting hard X-rays were identified (corresponding to the
known sources \object{[KLS2002] HX3 west} and 
\object{CXO J192318.5+140305} discovered with ASCA 
and \chandra, respectively), each of which has an additional point 
source inside and shows a power-law spectrum with $\Gamma\ \approx\ 1.8$.
Based on their X-ray emission, both sources can be classified as PWN 
candidates.} 
   {}


   \maketitle
%

\section{Introduction}\label{intro}

Supernova remnants (SNRs) are galactic objects that are formed
after a supernova (SN) explosion at the end of the life of a star. 
An SN explosion instantaneously releases energy and matter into the
ambient interstellar medium (ISM), carving out new structures inside
a galaxy and enriching it with elements that were formed inside 
the progenitor star and in the explosion. In the shock waves of SNRs,
particles can be accelerated and become Galactic cosmic rays. 
Multi-frequency studies of SNRs from radio to X-rays, in some cases
even to $\gamma$-rays, help to understand the physics of the
shock waves and the interaction between the SNR and the ISM.

The SNR W51C (G49.1--0.1) is a Galactic SNR in the W51 complex, an extended
radio source at the tangential point of the Sagittarius arm. The W51 complex
comprises two additional structures, W51A and W51B, which both harbour compact
\hii\ regions. High-velocity molecular gas that runs
almost parallel to the Galactic plane, known as the high-velocity stream
\citep{1979ApJ...232..451M},
has been detected on the western side of the SNR.
Located in an area with a large number of \hii\ regions and dense molecular
gas, W51C is a prime example of the interaction of an SNR blast
wave with the multi-phase interstellar medium in our Galaxy.

The radio continuum of SNR W51C shows a thick arc-like structure with an
extent of about 30\arcmin\ 
\citep{1991MNRAS.250..127C,1995MNRAS.275..755S}.
It is open to the north and its projected western edge coincides with W51B.
The distance of 6\,kpc to the SNR has been
estimated from molecular-line observations
\citep{1997ApJ...475..194K,1997ApJ...485..263K,1997AJ....114.2058G}
and places it behind the ridge of molecular gas at 5.6 kpc. 
Assuming a Sedov model, \citet{1995ApJ...447..211K} derived
an SNR age of $\sim 3 \times 10^{4}$~yr.

On its western side, where W51C merges with W51B in the radio image,
shocked atomic \hi\ was found \citep{1991ApJ...382..204K}.
Furthermore, \citet{1997ApJ...485..263K} found CO and HCO$^{+}$ emission,
probably arising from molecules destroyed by a fast dissociative shock
and reformed behind the shock front
\citep{1997ApJ...475..194K}.
Masers at 1720~MHz (OH) have also been found in W51C, between the SNR and the
high-velocity stream of molecular gas
\citep{1997AJ....114.2058G};
OH (1720~MHz) emission, unaccompanied by maser emission from
other OH ground-state transitions,
is thought to arise in cooling gas behind non-dissociative shocks
and is used to locate the site of shock-cloud
interactions.

\begin{figure}
\centering
\includegraphics[width=0.5\textwidth]{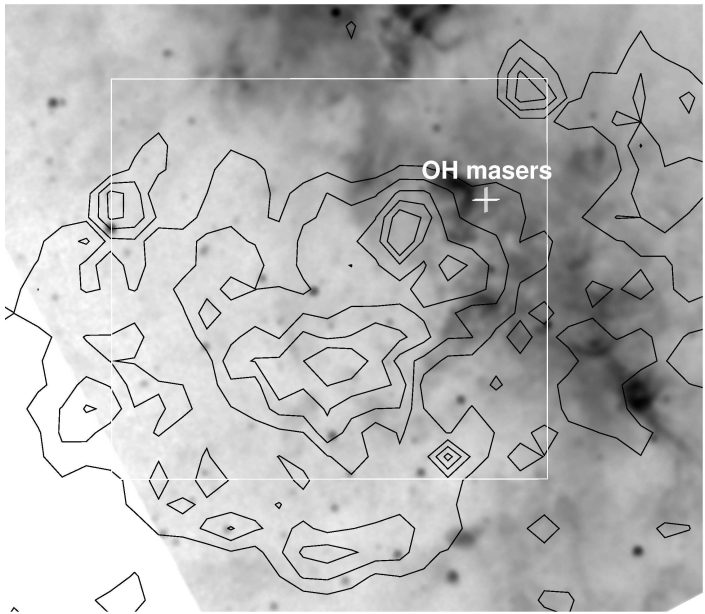}
\caption{\label{xir}
A Midcourse Space Experiment 
\citep[MSX,][]{2003.AFRL.VS.TR...E}
image in the infrared 
taken at 8.3 $\mu$m, together with ROSAT PSPC 
(0.1 -- 2.4 keV) contours. The MSX image shows 
the emission from interstellar dust.
The white cross marks the position of the OH (1720 MHz) masers.
The white box indicates the shown field of the sky in Fig.\,\ref{rgb}.
}
\end{figure}

Diffuse X-ray emission was observed from W51B and W51C with the {\it Einstein} 
Imaging Proportional Counter \citep{1990ApJS...73..781S} and the Position 
Sensitive Proportional Counter (PSPC) of the R\"ontgen Satellite (ROSAT)
mainly below 2~keV \citep{1995ApJ...447..211K}.
The X-ray shell of W51C in the east and the western diffuse X-ray emission
match the radio shell, while the emission gap observed in the soft X-rays
between the bright central region and the western part coincides with W51B
(see Fig.\,\ref{xir}).
Koo et al. (\citeyear{2002AJ....123.1629K}, hereafter KLS2002) 
derived a temperature of $\sim$0.3~keV from 
data taken with the Advanced Satellite for Cosmology and Astrophysics 
(ASCA) 
for the central soft SNR emission, assuming thermal emission in
collisional ionisation equilibrium.
The hard (2.6 -- 6.0 keV) X-ray image of ASCA
is very different from the ROSAT image with the brightest
regions seeming to be associated with compact \hii\ regions.

The remnant W51C is very likely the result of a core-collapse SN
and should have an associated compact object.
Images taken with the \chandra\ Advanced CCD Imaging Spectrometer in the soft 
band \citep[0.3 -- 2.1 keV,][]{2005ApJ...633..946K}
show several X-ray bright clumps.
The spectra of the diffuse emission suggest an enhanced abundance of sulfur.
In the hard band (2.1 -- 10.0 keV), the point-like source 
CXO J192318.5+140305 was found, which is 
surrounded by an extended emission, making it a likely candidate for a pulsar 
wind nebula (PWN). 
This structure is similar to PWNe associated with several other SNRs, 
which all have a $\sim 1$~pc diameter
bright core with a central pulsar or a point source. There is an indication of
an unusual hardening at the ends of the envelope of this source in the
\chandra\ data, but uncertainties are large. Its spectrum can be fitted
with a power-law model \citep[$\Gamma=1.8\pm0.3$,][]{2005ApJ...633..946K}.
Recently, \citet{2013PASJ...65...42H} presented the results of a 
study of the inner parts of the SNR W51C performed with the \suzaku\ X-ray 
Imaging Spectrometer (XIS).

At higher energies $\gamma$-ray emission from the W51C region has been 
detected with the Large Area Telescope (LAT) on board {\it Fermi}
\citep{2009ApJ...706L...1A}, the High Energy Stereoscopic System 
(H.E.S.S.)\footnote{Reported by Fiasson et al.\ (2009,
Proc., 31th ICRC).}, and the MAGIC telescopes
\citep{2012A&A...541A..13A}.
The GeV emission observed with {\it Fermi} is extended and seems to be consistent
with the radio and X-ray extent of the remnant.

In this paper we present the analysis of the central part of the SNR W51C
from a deep observation with the X-ray Multi-Mirror Mission 
\citep[\xmm,][]{2001A&A...365L...1J}.  

\section{Data}

The SNR W51C was observed with \xmm\ on April 8, 2009, (obs.\,ID 0554690101)
with the European Photon Imaging Cameras
\citep[EPICs,][]{2001A&A...365L..18S,2001A&A...365L..27T} as
prime instruments using the medium filters. We used
HEASOFT ver.\,6.12 and the \xmm\ Science Analysis System ver.\,12.0.0 to
analyse the data.

Since the emission from the SNR W51C fills almost the entire field of view 
(FOV) of 
the EPICs, we used the \xmm\ Extended Source Analysis Software 
(ESAS)\footnote{\it http://heasarc.nasa.gov/docs/xmm/xmmhp\_xmmesas.html}
to analyse the data. Using ESAS, we filtered out bad time intervals
caused by soft proton flares, resulting in good exposures of
40, 51, and 54 ks for EPIC-pn, MOS1, and MOS2, respectively.
In addition, we checked whether any CCD of the MOS detectors was in an 
anomalous state. Since CCD4 of MOS1 showed an enhancement in the background
below 1 keV and was therefore in an anomalous 
state\footnote{XMM-ESAS Users Guide, Snowden \& Kuntz 2012},
this CCD was excluded from further analysis. 
The filtered events were used to perform source detection and create 
data with events of the extended
emission. For further analysis, we selected single and double pixel
events (PATTERN = 0 to 4) for the EPIC-pn and single to quadruple-pixel 
events (PATTERN = 0 to 12) for the EPIC-MOS1 and MOS2.
To remove the detector background, data taken with the filter wheel
closed (FWC) were binned to images and subtracted from the images
of the observation after rescaling the FWC image using the count rates
in the corners of the detectors, which are not exposed to the sky.

We are primarily interested in the extended diffuse emission of
the SNR. Therefore, we ran the ESAS routines {\it cheese} 
(or {\it cheese-bands} for processing multiple energy bands) to detect 
point sources, which can then be removed in the following steps.
These routines run source detection and create lists and masks of the 
detected point sources.
The masks are used to remove the point sources from the images
and the list of detected sources is used to exclude the sources in
the extraction regions for spectral analysis.
For the source detection we used 
a flux threshold of $10^{14}$ erg cm$^{-2}$ s$^{-1}$, a S/N 
ratio of 2, and a minimum separation of 40\arcsec\ for the point sources.
In total, 48 point sources were found in the FOV, which we excluded 
when creating images and extracting spectra. We only checked
the spectra and counterparts of the potential pulsar candidates,
i.e.\ hard X-ray sources with associated hard extended emission around
them (see Sect.\,\ref{point}).

\begin{figure}
\centering
\includegraphics[width=0.5\textwidth]{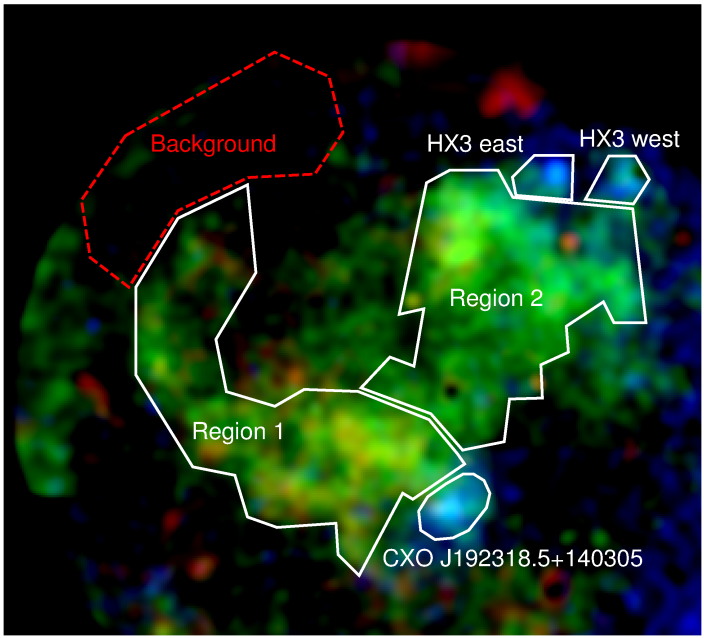}
\caption{\label{rgb}
Three-colour image of the mosaic images (red: 0.3 -- 1.0 keV, green:
1.0 -- 2.0 keV, blue: 2.0 -- 8.0 keV). Extraction regions used for the 
spectral analysis are shown. Point sources are removed.
}
\end{figure}

To obtain deep images of the SNR in different energy bands,
the images of the EPIC-pn and MOS1 and 2 were merged and divided by 
exposure maps, which take the vignetting effects into account.
Figure \ref{rgb} shows a three-colour presentation of the
images in the bands 0.3 -- 1.0 keV (red), 1.0 -- 2.0 keV (green), and  
2.0 -- 8.0 keV (blue). These images, which are shown in 
logarithmic scale, were adaptively smoothed
and then smoothed again using a Gaussian kernel of the size of 5 pixels
for a better presentation of the faint diffuse emission.
The holes in the images resulting from excluding point sources are not 
filled.
The mosaicing of the three EPICs and the adaptive smoothing 
even out most of the holes as well as the chip gaps as can be seen in 
Fig.\,\ref{rgb}.

Using these images,
extraction regions for the spectral analysis were defined
based on surface brightness and X-ray colour (for details see Sect.\,\ref{spec}). 
The spectra were rebinned with a minimum of 50 counts per bin.
For a comparison of the distribution of the X-ray emission and the surrounding 
colder medium, we also downloaded an image from the Midcourse Space Experiment 
\citep[MSX, see Fig.\ref{xir},][]{2003.AFRL.VS.TR...E}.
To estimate the local X-ray background, a background extraction region was
defined north of the SNR emission. The fainter regions in the east to south-east
or in the west are not appropriate as this is where the blast wave emission is 
expected. The background spectra were scaled to the source spectra using the areas 
of the extraction regions, taking into account that CCD4 in anomalous 
state and CCD6, which was damaged after being hit by a micrometeorite, are 
excluded for MOS1.

\section{Spectral analysis}\label{spec}

The spectra were analysed using the X-ray spectral fitting package XSPEC
Ver.\ 12.8.1 and the atomic data base AtomDB 2.0.2\footnote{\it
http://www.atomdb.org}.
The emission of the SNR fills most of the FOV of the analysed 
observation. 
We divided the data into regions that are likely to have emission of similar 
nature based on the images in the broad band and the 
three-colour image (Fig.\,\ref{rgb}).
Unfortunately, the statistics of the data were not high enough to analyse
small regions. We were also not able to extract meaningful spectra of the 
fainter regions that correspond to the outer shock of the SNR, for example.
We therefore focussed the spectral analysis on the inner X-ray bright region,
which we divided into two, one for the arc-like structure in the south-east and
one covering the region north-west of the arc region located next to
a band of molecular gas showing dust emission and harbouring compact \hii\ regions
(Fig.\,\ref{xir}).
The two regions are to a certain extent consistent with the extraction regions 
1 and 2 used by \citet{2013PASJ...65...42H} for the analysis of \suzaku\ XIS data
and so we also call them regions 1 and 2.
These two extraction regions also correspond more or less to the regions
XS and XN in \citet{2002AJ....123.1629K}.

In addition, we extracted spectra for the diffuse emission around
two hard point-like sources (one at the position of CXO J192318.5+140305 
and one north of the X-ray bright region, corresponding to sources 
[KLS2002] HX2 and HX3 west, respectively),
as well as a region covering \object{[KLS2002] HX3 east}, 
which has been resolved into two 
sources in the \xmm\ data (see Sect.\,\ref{point}).

The observed X-ray spectrum includes the particle background measured by
the detector and the X-ray background in addition to the actual source emission.
To subtract the particle background, we extracted spectra from the
FWC data at the same detector positions as for the extraction regions of the
analysed source data. Since the closed filter wheel blocks all X-ray emission
from outside and the soft protons from solar flares, this background accounts 
for the contamination by cosmic rays. 
These FWC-background spectra are subtracted as backgrounds in XSPEC. 
The FWC-background subtracted spectra still show some flourescent lines 
coming from the instruments, which can vary from observation to observation and 
so cannot be completely removed by subtracting the FWC data. In addition,
the spectra include the X-ray background.
The time intervals with enhanced soft proton background had already been 
filtered out in the beginning of the data analysis. The source spectra showed 
no signs of residual particle-background continuum, which should otherwise have
been modelled as an additional unfolded power-law 
component, not affected by detector response. 
The X-ray background was estimated by modelling the source spectrum and
the local background spectrum extracted outside the SNR simultaneously.
In doing so, spectra of
all detectors were fitted simultaneously for each source region. 
The background spectra were modelled with a spectrum
consisting of a component for the thermal emission from the Local Bubble, 
one for the thermal emission from the Galactic halo, and a power-law describing 
the extragalactic X-ray background.  All the parameters of the
background components were linked for all spectra. The assumed temperatures for
the thermal X-ray background and the photon index for the extragalactic 
background were fixed to values suggested by Snowden et al.\ 
(\citeyear{2008A&A...478..615S}, see also the XMM-ESAS Users Guide).
Another process that can also create additional emission lines
in the observed spectra is the solar wind charge exchange 
\citep[SWCX,][]{2004ApJ...610.1182S}. Therefore, in total four Gaussians have 
been included to model possible SWCX emission and the fluorescent lines formed 
in the detector.
The parameters of the Gaussians were free, as the centroids and the
line fluxes can vary for the fluorescent lines between detectors and also 
between different positions on one detector.

\subsection{Diffuse emission}

\begin{table*}
\caption{
\label{vneitab}
Fit parameters for the SNR emission.
}
\centering
\begin{tabular}{lccccccc}
\hline\hline\
ID & $N_{\rm H}$ & $kT$ & $\tau$ & Ne & Mg & red.\ $\chi^2$ & DOF$^\ast$\\
& [$10^{22}$ cm$^{-2}$] & [keV] & [$10^{11}$ s cm$^{-3}$] & $\times$ solar & $\times$ solar & & \\
\hline
Region 1 & 1.6 & 0.68 & 1.1 & 1.0$^\star$ & 1.0$^\star$ & 1.3 & 1034\\
 & (1.5 -- 1.7) & (0.66 -- 0.74) & (0.8 -- 1.5) & & & & \\
Region 2 & 2.0 & 0.59 & 8.1 & 3.7 & 2.0 & 1.2 & 970 \\
 & (1.9 -- 2.1) & (0.56 -- 0.61) & (6.2 -- 11.) & (3.5 -- 4.2) & (1.9 -- 2.2) & & \\
\hline
\end{tabular}
\vspace{-2mm}
\tablefoot{
$^\ast$Degrees of freedom. $^\star$Fixed to solar values like all the
other abundances.
Errors are 90\% confidence limits.
}
\end{table*}

\begin{figure*}
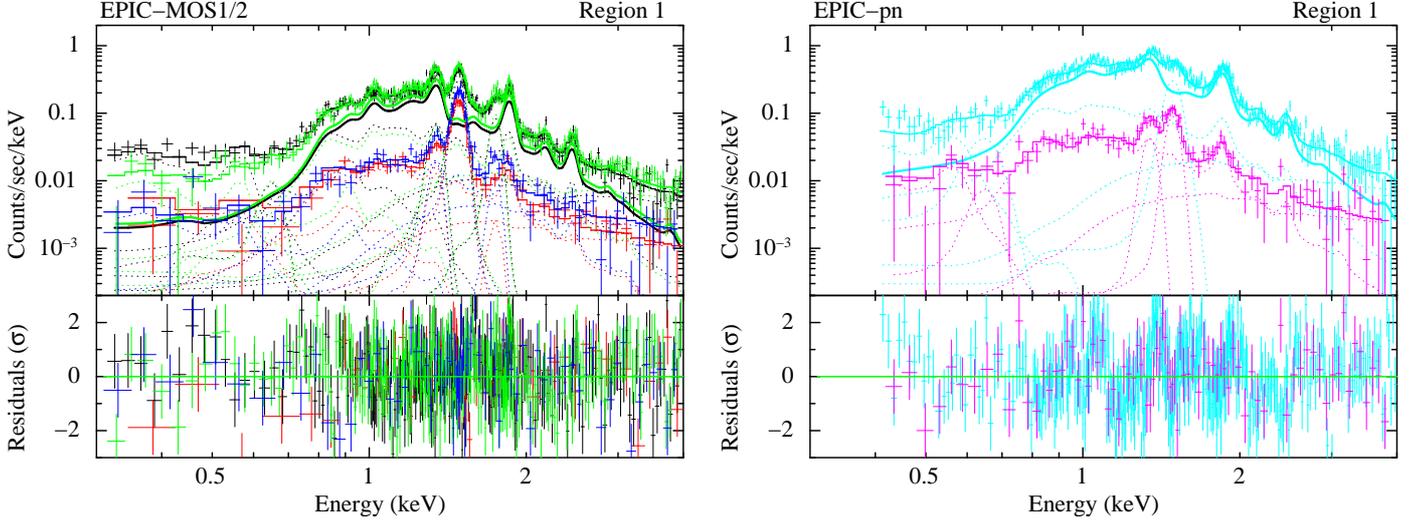

\centering
\includegraphics[height=0.49\textwidth,angle=270,bb=47 38 563 709,clip]{regBC_MS_mos.ps}
\hfill
\includegraphics[height=0.49\textwidth,angle=270,bb=47 38 563 709,clip]{regBC_MS_pn.ps}
\caption{\label{BCvnei}
EPIC spectra of region 1 (MOS1: black, MOS2: green, pn: cyan) and the
background region (MOS1: red, MOS2: blue, pn: magenta) and the best-fit
model. The thick solid lines show the source emission component (VNEI).
The dotted lines show all the other components included in the model. 
}
\end{figure*}

\begin{figure*}
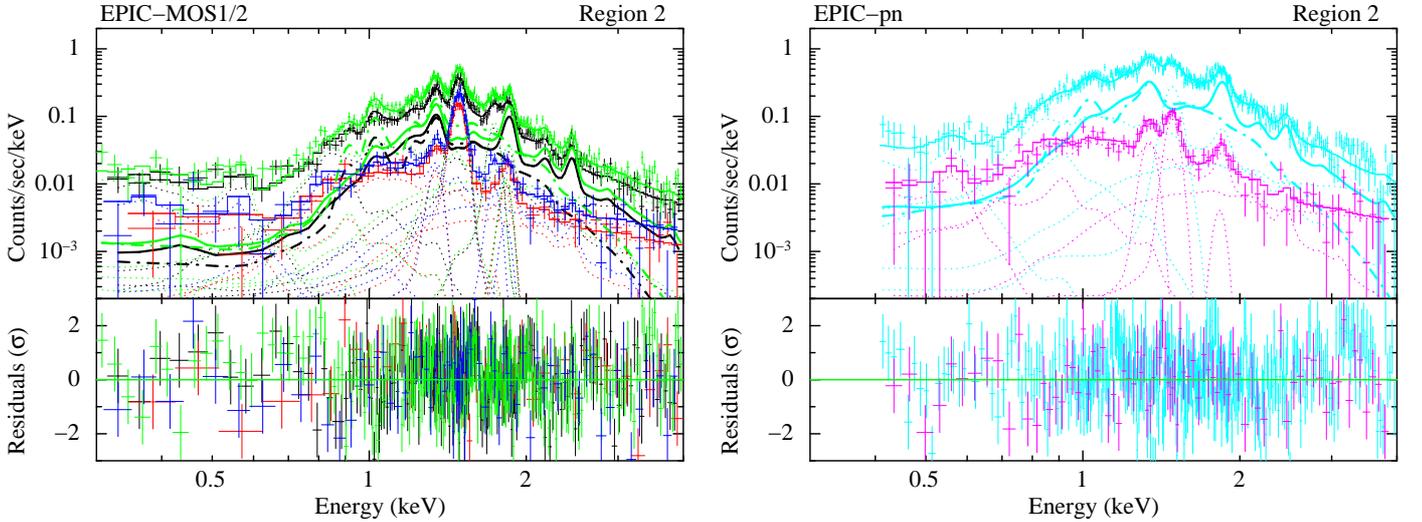

\centering
\includegraphics[height=0.49\textwidth,angle=270,bb=47 38 563 709,clip]{regEF_2vnei_mos.ps}
\hfill
\includegraphics[height=0.49\textwidth,angle=270,bb=47 38 563 709,clip]{regEF_2vnei_pn.ps}
\caption{\label{EF2vnei}
EPIC spectra of region 2 (MOS1: black, MOS2: green, pn: cyan) and the
background region (MOS1: red, MOS2: blue, pn: magenta) and the best-fit
model, assuming two VNEI components for the source emission 
(see Sect.\,\ref{ejecta}). 
The thick solid lines show the source emission component corresponding
to shocked ISM (VNEI with solar abundances, parameters fixed to the results
of region 1). The thick dash-dotted lines show the ejecta component.
The dotted lines show all the other components included in the model.
}
\end{figure*}

To model the emission of the SNR, an additional thermal emission component
was included for the source spectra. Best fit was achieved with a 
non-equilibrium
ionisation model in both regions. We used the model VNEI which allowed us
to vary the abundances. We used the solar element abundances determined by
\citet{1989GeCoA..53..197A} for all fits.
The best-fit parameter values are listed in Table 
\ref{vneitab}.

The spectrum of region 1 is well fitted with a VNEI model assuming solar
abundances (Fig.\,\ref{BCvnei}). 
If we also assume solar abundances for all elements for
region 2, the fit is not sufficiently good and there are residuals especially 
at energies $>$1 keV. The fit improves if we free the abundances for Ne 
and Mg. The best fit indicates an overabundance of these two elements
(see Table \ref{vneitab}).

The X-ray emission from region 1 is consistent with emission 
from shocked ISM. 
This region 1 has a much higher surface brightness than the outer shock region 
to the east, which is seen in the radio map and is very faint in the X-rays
(see the ROSAT contours in Fig.\,\ref{xir}). The interior bright emission
indicates that there was a region with higher density around the SN, i.e.\
circumstellar matter ejected from the progenitor. Inside this circumstellar
matter, in region 2
we find enhanced abundances of selected elements (Ne, Mg), indicating
emission from stellar ejecta.

\subsection{Point-like sources}\label{point}

There are three bright hard sources clearly visible in Fig.\,\ref{rgb}, 
one of which coincides with the source CXO J192318.5+140305 suggested to be
a PWN by \citet{2005ApJ...633..946K}. The two hard sources at the northern rim
of the FOV of the \xmm\ observation
correspond to the ASCA sources [KLS2002] HX3 east 
and west. We extracted and analysed the spectra of
the diffuse emission in these regions.

\begin{table*}
\caption{
\label{powtab}
Fit parameters for the hard extended emission.
}
\centering
\begin{tabular}{lcccccc}
\hline\hline\
ID & RA, Dec$^\dagger$ & $N_{\rm H}$ & Photon Index & unabsorbed Flux & red.\ $\chi^2$ & DOF$^\ast$\\
& (2000.0) & [$10^{22}$ cm$^{-2}$]  &  & (0.3 -- 10 keV) & & \\
 & & & & [$10^{-13}$ erg s$^{-1}$ cm$^{-2}$] & & \\ 
\hline
CXO J192318.5+140305 & 19 23 18.6, +14 03 03 & 1.00 & 1.89 & 9.7 & 1.3 & 390\\
& & (0.88 -- 1.13) & (1.75 -- 2.03) & (8.2 -- 12.) & & \\
HX3 west & 19 22 48.4, +14 16 27 & 1.79 & 1.75 & 9.0 & 1.3 & 302 \\
& & (1.70 -- 2.51) & (1.67 -- 1.94) & (8.5 --16.) & & \\
\hline
\end{tabular}
\vspace{-2mm}
\tablefoot{
$^\dagger$Positions of the point sources as determined by the source 
detection routine of ESAS.
$^\ast$Degrees of freedom.
Errors are 90\% confidence limits.
}
\end{table*}

The source HX3 east has a counterpart in the radio 
(G49.2--0.3), which has been identified as a compact \hii\ region 
\citep{1997ApJ...475..194K}
and has thus been classified as emission from massive stars in G49.2--0.3.
In the presented \xmm\ data, the hard emission at the position of HX3 east
has been resolved into two sources that are 17\arcsec\ apart 
(RA, Dec = 19 23 01.7, +14 16 34 and 19 23 01.8, +14 16 17) and have similar 
X-ray colours. We extracted the spectrum of a region that includes both sources.
The spectrum was first fitted with a power-law model for the source emission.
The background was modelled including the background components as described
before.
The resulting photon index is $\Gamma$ = 
3.2 (2.6 -- 3.8)\footnote{All errors in brackets are 90\% confidence intervals 
throughout the paper.}
and the intrinsic absorption is
\nh = $3.5 (3.0 - 4.9) \times\ 10^{22}$ cm$^{-2}$ (reduced $\chi^2$ = 1.4 for 299 
degrees of freedom). A thermal emission model for a plasma in 
non-equilibrium ionisation instead of the non-thermal model improved the fit
slightly (reduced $\chi^2$ = 1.3 for 297 degrees of freedom) and resulted
in an intrinsic absorption of \nh = $3.9 (3.5 - 4.3) \times\ 10^{22}$ cm$^{-2}$,
a temperature of $kT = 2.4 (1.8 - 3.5)$ keV, and an ionisation timescale of
$\tau = 2.0 (1.2 - 3.3) \times 10^{10}$ s cm$^{-3}$. The high photon index 
in the non-thermal fit is indicative of a thermal emission, whereas the
high column density obtained in both fits suggests an emission origin embedded
in a dense medium. The temperature and the ionisation timescale are also 
consistent with winds of young massive stars, confirming the results of
\citet{2002AJ....123.1629K}.

The source spectra of CXO J192318.5+140305 and HX3 west
were modelled with a power-law component. The results are
summarised in Table \ref{powtab}. Both sources are well fitted with
a power-law spectrum with a photon index in the range of $\Gamma = 1.7 - 2.0$.

\section{Discussion}

\subsection{Ejecta mass}\label{ejecta}

The well-defined emission from region 2 allows us to estimate the 
amount of ejecta. 
To do so, we fitted the spectra of region 2 with a combined source model
consisting of two VNEI models. For the first VNEI component, the parameters 
$kT$ and $\tau$ were fixed
to the best-fit values of region 1. The abundances were all set to solar.
For the second VNEI spectrum, the parameters $kT$, $\tau$, and the abundances
of Ne and Mg were free, while all other abundances were set to zero first.
In the subsequent steps we also fitted the abundances of all other elements
by additionally freeing them one by one, but no other element showed a
significantly enhanced abundance.
This spectral component will allow us to estimate the amount of Ne and Mg
in the ejecta. The spectra of region 2 with the best-fit model is shown in 
Fig.\,\ref{EF2vnei}. As the best-fit parameters, we get the following
values with lower 90\% confidence limits for the normalisation
of $norm = \frac{10^{-14}}{4 \pi D^{2}} \times \int n_{\rm H}~n_{e}~dV = 
1.4 (> 0.7) \times 10^{-3}$ cm$^{-5}$,
with $D$ being the distance to the source, and the abundances for
Ne = 180 ($>$ 10) and Mg = 50 ($>$ 20) $\times$ solar values. 
Since the fit parameters for the normalisation and the abundances are 
correlated to each 
other in this model, which is dominated by emission lines, it is not 
possible to constrain the upper confidence limits for these values.

At X-ray emitting temperatures, we can assume that all elements are
almost fully ionised. In this case, we get $n_{e} = 1.21 n_{\rm H}$ for the 
relation between the electron and H densities.
The extraction region has an extent of about 10\arcmin.
For the volume of the emitting region, we take the size of the extraction
region and assume a depth of 10\arcmin, corresponding to the extent of the 
projected volume.
For the calculations, we assume a distance of $D = 6$ kpc.
We thus get a particle number density of 
$n_{\rm H} = 0.07~(>0.05)$ cm$^{-3}$. 
For this density, we derive an ionisation time of
$\tau/(1.21n_{\rm H}) = 3.0 \times 10^{5}$ yr
using the ionisation timescale obtained from the spectral fit of 
region 2. This time is one order of magnitude higher than the
estimated age of the SNR  \citep[][see Sect.\,\ref{intro}]
{1995ApJ...447..211K}. 
The ionisation timescale gives a characteristic
number for the X-ray emitting plasma after it has been shocked and heated, 
and does not thus necessarily represent the real age of the SNR. As can be
verified in Table \ref{vneitab}, the ionisation timescale $\tau$ is 
significantly higher for region 2 than for the region 1, which is located 
outside region 2. This high ionisation age in the likely centre of the
SNR surrounded by shocked circumstellar medium might indicate that the 
plasma is overionised, as observed in some mixed-morphology SNRs in our 
Galaxy \citep[][and references therein]{2013ApJ...777..145L}.

Using the abundances obtained from the spectral fit, we estimate the 
masses of
$M_{\rm Ne} = 2.9~(>0.16)~M_{\sun}$ and 
$M_{\rm Mg} = 0.30~(>0.12)~M_{\sun}$ for the ejecta. Because of the large
uncertainties in determining the volume as well as the spectral fit parameters,
these numbers only give a crude estimate of the masses. 
The comparison to calculated nucleosynthesis yields of SNe Type II by
\citet{1997NuPhA.616...79N} has shown that these mass limits for Ne and Mg 
point to a very massive progenitor with a mass of $> 20 M_{\sun}$.

\subsection{Pulsar wind nebula candidates}

The spectral analysis of the regions around CXO J192318.5+140305 and HX3 west
have shown that both sources have a power-law spectrum with photon indices
that are comparable and consitent with a PWN. The flux and the extent
of the diffuse emission are also similar.

The hard source at the position of HX3 west consists of an extended emission 
and a hard point source, which is unfortunately located at the rim of the FOV. 
This makes a detailed analysis of the point source
impossible. The three-colour \chandra\ image presented by 
Koo et al.\ (\citeyear{2005ApJ...633..946K}, their Fig.\,1) 
also shows a blue (thus hard) point source at the north-western
rim of the extended hard emission. As noted before,
the hard sources at the positions of CXO J192318.5+140305 and  
HX3 west are very similar, both consisting of a diffuse emission plus a 
point source and the photon index of the extended emission being consistent 
with a PWN.
The measured fluxes of the extended emission correspond to luminosities of
$L_{\rm 0.3 - 10.0 keV} = 4.2 (3.5 - 52.) \times 10^{33}$ erg s$^{-1}$ 
and $3.9 (3.7 - 69.) \times 10^{33}$ erg s$^{-1}$ for
CXO J192318.5+140305 and HX3 west, respectively,
if we assume a distance of $D = 6$ kpc.
At both positions, faint radio sources were found 
\citep{1998AJ....115.1693C,2002AJ....123.1629K}; 
therefore, we conclude that both sources are likely candidates
of a PWN associated to the SNR W51C and require further investigation.

\section{Summary}

We have studied the Galactic SNR W51C using an observation of its 
central part carried out with \xmm. It was not possible to analyse 
the X-ray emission of the outer blast wave as it was not covered 
completely by the pointing and the observed emission was too faint.
However, the spectral analysis of the inner regions revealed that there
are two distinct emission regions inside the SNR. 

The first region
has an arc-like morphology, more or less aligned to the morphology of 
the blast wave outside, but is much brighter. Its spectrum is well 
reproduced by thermal emission from plasma with non-equilibrium
ionisation. The element abundances are consistent with solar values,
indicating that the origin of this emission is shocked ambient medium.
The shape and the higher X-ray brightness of this region suggest that
the shocked material is the circumstellar medium of the progenitor.

The diffuse X-ray emission from the second region, located farther inside 
and in projection next to the region filled with colder gas and dust,  
is more highly absorbed and seems to arise from plasma close to
collisional ionisation equilibium. Significantly enhanced abundances
are measured for Ne and Mg. From the abundances
and the density derived from the spectra, we calculated the masses
of Ne and Mg in the ejecta and conclude that the progenitor
was a massive star with a mass comparable to or higher than $20 M_{\sun}$.

In addition, we analysed the diffuse X-ray emission detected around
two hard point sources, one of which has already been studied in 
detail by \citet{2005ApJ...633..946K} using \chandra\ data. The 
second hard source is located north-west of the region, close to 
the OH (1720~MHz) masers that have been found in W51C.
The extended emission of both regions is well modelled with an 
absorbed power-law spectrum with a photon index of 
$\Gamma \approx\ 1.8$. The luminosities are 
$L_{\rm 0.3 - 10.0 keV} \approx\ 4 \times 10^{33}$ 
erg s$^{-1}$ for both regions; the value is rather low, but still
consistent with values measured for older PWNe 
\citep[e.g.,][]{2010AIPC.1248...25K}. In addition, the morphology of both 
sources consisting of a diffuse emission with an extent of $\sim$1\arcmin\
and a point source supports their classification as PWN candidates.

The \xmm\ observation has enabled us to study the diffuse emission 
in the SNR W51C in detail and helped us to make a step forward in the
classification of the underlying components of the complex X-ray emission 
of the SNR. Further observations will be necessary to study the
blast wave of the SNR and to identify the possibly associated pulsar/PWN.

\begin{acknowledgements}
M.S.\ and G.W.\ acknowledge support by the Deutsche Forschungsgemeinschaft 
through the Emmy Noether Research Grant SA 2131/1-1.
This research was funded through the BMWi/DLR grant 50 OR 1009.
This research made use of data products from the Midcourse Space Experiment. 
Processing of the data was funded by the Ballistic Missile Defense Organization 
with additional support from NASA Office of Space Science. 
This research has also made use of the NASA/ IPAC Infrared Science Archive, 
which is operated by the Jet Propulsion Laboratory, California Institute of 
Technology, under contract with the National Aeronautics and Space Administration.
\end{acknowledgements}

\bibliographystyle{aa} 
\bibliography{../../bibtex/w51c,../../bibtex/xraytel,../../bibtex/my,../../bibtex/sne,../../bibtex/snrs}

\begin{thebibliography}{25}
\expandafter\ifx\csname natexlab\endcsname\relax\def\natexlab#1{#1}\fi

\bibitem[{{Abdo} {et~al.}(2009){Abdo}, {Ackermann}, {Ajello}, {Baldini},
  {Ballet}, {Barbiellini}, {Baring}, {Bastieri}, {Baughman}, {Bechtol},
  {Bellazzini}, {Berenji}, {Blandford}, {Bloom}, {Bonamente}, {Borgland},
  {Bouvier}, {Bregeon}, {Brez}, {Brigida}, {Bruel}, {Burnett}, {Buson},
  {Caliandro}, {Cameron}, {Caraveo}, {Casandjian}, {Cecchi}, {{\c C}elik},
  {Chekhtman}, {Cheung}, {Chiang}, {Ciprini}, {Claus}, {Cohen-Tanugi},
  {Cominsky}, {Conrad}, {Cutini}, {Dermer}, {de Angelis}, {de Palma}, {Digel},
  {Dormody}, {Silva}, {Drell}, {Dubois}, {Dumora}, {Farnier}, {Favuzzi},
  {Fegan}, {Focke}, {Fortin}, {Frailis}, {Fukazawa}, {Funk}, {Fusco},
  {Gargano}, {Gasparrini}, {Gehrels}, {Germani}, {Giavitto}, {Giebels},
  {Giglietto}, {Giordano}, {Glanzman}, {Godfrey}, {Grenier}, {Grondin},
  {Grove}, {Guillemot}, {Guiriec}, {Hanabata}, {Harding}, {Hayashida}, {Hays},
  {Hughes}, {Jackson}, {J{\'o}hannesson}, {Johnson}, {Johnson}, {Johnson},
  {Kamae}, {Katagiri}, {Kataoka}, {Katsuta}, {Kawai}, {Kerr}, {Kn{\"o}dlseder},
  {Kocian}, {Kuss}, {Lande}, {Latronico}, {Lemoine-Goumard}, {Longo},
  {Loparco}, {Lott}, {Lovellette}, {Lubrano}, {Makeev}, {Mazziotta}, {McEnery},
  {Meurer}, {Michelson}, {Mitthumsiri}, {Mizuno}, {Moiseev}, {Monte},
  {Monzani}, {Morselli}, {Moskalenko}, {Murgia}, {Nakamori}, {Nolan}, {Norris},
  {Nuss}, {Ohsugi}, {Okumura}, {Omodei}, {Orlando}, {Ormes}, {Paneque},
  {Parent}, {Pelassa}, {Pepe}, {Pesce-Rollins}, {Piron}, {Porter}, {Rain{\`o}},
  {Rando}, {Razzano}, {Reimer}, {Reimer}, {Reposeur}, {Ritz}, {Rodriguez},
  {Romani}, {Roth}, {Ryde}, {Sadrozinski}, {Sanchez}, {Sander}, {Saz
  Parkinson}, {Scargle}, {Schalk}, {Sgr{\`o}}, {Siskind}, {Smith}, {Smith},
  {Spandre}, {Spinelli}, {Strickman}, {Suson}, {Tajima}, {Takahashi},
  {Takahashi}, {Tanaka}, {Thayer}, {Thayer}, {Thompson}, {Tibaldo}, {Tibolla},
  {Torres}, {Tosti}, {Tramacere}, {Uchiyama}, {Usher}, {Vasileiou}, {Venter},
  {Vilchez}, {Vitale}, {Waite}, {Wang}, {Winer}, {Wood}, {Yamazaki}, {Ylinen},
  \& {Ziegler}}]{2009ApJ...706L...1A}
{Abdo}, A.~A., {Ackermann}, M., {Ajello}, M., {et~al.} 2009, \apjl, 706, L1

\bibitem[{{Aleksi{\'c}} {et~al.}(2012){Aleksi{\'c}}, {Alvarez}, {Antonelli},
  {Antoranz}, {Asensio}, {Backes}, {Barres de Almeida}, {Barrio}, {Bastieri},
  {Becerra Gonz{\'a}lez}, {Bednarek}, {Berger}, {Bernardini}, {Biland},
  {Blanch}, {Bock}, {Boller}, {Bonnoli}, {Borla Tridon}, {Bretz},
  {Ca{\~n}ellas}, {Carmona}, {Carosi}, {Colin}, {Colombo}, {Contreras},
  {Cortina}, {Cossio}, {Covino}, {Da Vela}, {Dazzi}, {De Angelis}, {De Caneva},
  {De Cea del Pozo}, {De Lotto}, {Delgado Mendez}, {Diago Ortega}, {Doert},
  {Dom{\'{\i}}nguez}, {Dominis Prester}, {Dorner}, {Doro}, {Eisenacher},
  {Elsaesser}, {Ferenc}, {Fonseca}, {Font}, {Fruck}, {Garc{\'{\i}}a L{\'o}pez},
  {Garczarczyk}, {Garrido}, {Giavitto}, {Godinovi{\'c}}, {Gonz{\'a}lez
  Mu{\~n}oz}, {Gozzini}, {Hadasch}, {H{\"a}fner}, {Herrero}, {Hildebrand},
  {Hose}, {Hrupec}, {Huber}, {Jankowski}, {Jogler}, {Kadenius}, {Kellermann},
  {Klepser}, {Kr{\"a}henb{\"u}hl}, {Krause}, {La Barbera}, {Lelas}, {Leonardo},
  {Lewandowska}, {Lindfors}, {Lombardi}, {L{\'o}pez}, {L{\'o}pez-Coto},
  {L{\'o}pez-Oramas}, {Lorenz}, {Makariev}, {Maneva}, {Mankuzhiyil},
  {Mannheim}, {Maraschi}, {Mariotti}, {Mart{\'{\i}}nez}, {Mazin}, {Meucci},
  {Miranda}, {Mirzoyan}, {Mold{\'o}n}, {Moralejo}, {Munar-Adrover},
  {Niedzwiecki}, {Nieto}, {Nilsson}, {Nowak}, {Orito}, {Paiano}, {Paneque},
  {Paoletti}, {Pardo}, {Paredes}, {Partini}, {Perez-Torres}, {Persic}, {Pilia},
  {Pochon}, {Prada}, {Prada Moroni}, {Prandini}, {Puerto Gimenez}, {Puljak},
  {Reichardt}, {Reinthal}, {Rhode}, {Rib{\'o}}, {Rico}, {R{\"u}gamer},
  {Saggion}, {Saito}, {Saito}, {Salvati}, {Satalecka}, {Scalzotto}, {Scapin},
  {Schultz}, {Schweizer}, {Shore}, {Sillanp{\"a}{\"a}}, {Sitarek}, {Snidaric},
  {Sobczynska}, {Spanier}, {Spiro}, {Stamatescu}, {Stamerra}, {Steinke},
  {Storz}, {Strah}, {Sun}, {Suri{\'c}}, {Takalo}, {Takami}, {Tavecchio},
  {Temnikov}, {Terzi{\'c}}, {Tescaro}, {Teshima}, {Tibolla}, {Torres},
  {Treves}, {Uellenbeck}, {Vogler}, {Wagner}, {Weitzel}, {Zabalza}, {Zandanel},
  \& {Zanin}}]{2012A&A...541A..13A}
{Aleksi{\'c}}, J., {Alvarez}, E.~A., {Antonelli}, L.~A., {et~al.} 2012, \aap,
  541, A13

\bibitem[{{Anders} \& {Grevesse}(1989)}]{1989GeCoA..53..197A}
{Anders}, E. \& {Grevesse}, N. 1989, {\rm Geochimica et Cosmochimica Acta}, 53,
  197

\bibitem[{{Condon} {et~al.}(1998){Condon}, {Cotton}, {Greisen}, {Yin},
  {Perley}, {Taylor}, \& {Broderick}}]{1998AJ....115.1693C}
{Condon}, J.~J., {Cotton}, W.~D., {Greisen}, E.~W., {et~al.} 1998, \aj, 115,
  1693

\bibitem[{{Copetti} \& {Schmidt}(1991)}]{1991MNRAS.250..127C}
{Copetti}, M.~V.~F. \& {Schmidt}, A.~A. 1991, \mnras, 250, 127

\bibitem[{{Egan} {et~al.}(2003){Egan}, {Price}, {Kraemer}, {Mizuno}, {Carey},
  {Wright}, {Engelke}, {Cohen}, \& {Gugliotti}}]{2003.AFRL.VS.TR...E}
{Egan}, M.~P., {Price}, S.~D., {Kraemer}, K.~E., {et~al.} 2003, Air Force
  Research Laboratory Technical Report AFRL-VS-TR-2003-1589

\bibitem[{{Green} {et~al.}(1997){Green}, {Frail}, {Goss}, \&
  {Otrupcek}}]{1997AJ....114.2058G}
{Green}, A.~J., {Frail}, D.~A., {Goss}, W.~M., \& {Otrupcek}, R. 1997, \aj,
  114, 2058

\bibitem[{{Hanabata} {et~al.}(2013){Hanabata}, {Sawada}, {Katagiri}, {Bamba},
  \& {Fukazawa}}]{2013PASJ...65...42H}
{Hanabata}, Y., {Sawada}, M., {Katagiri}, H., {Bamba}, A., \& {Fukazawa}, Y.
  2013, \pasj, 65, 42

\bibitem[{{Jansen} {et~al.}(2001){Jansen}, {Lumb}, {Altieri}, {Clavel}, {Ehle},
  {Erd}, {Gabriel}, {Guainazzi}, {Gondoin}, {Much}, {Munoz}, {Santos},
  {Schartel}, {Texier}, \& {Vacanti}}]{2001A&A...365L...1J}
{Jansen}, F., {Lumb}, D., {Altieri}, B., {et~al.} 2001, \aap, 365, L1

\bibitem[{{Kargaltsev} \& {Pavlov}(2010)}]{2010AIPC.1248...25K}
{Kargaltsev}, O. \& {Pavlov}, G.~G. 2010, X-ray Astronomy 2009; Present Status,
  Multi-Wavelength Approach and Future Perspectives, 1248, 25

\bibitem[{{Koo} \& {Heiles}(1991)}]{1991ApJ...382..204K}
{Koo}, B.-C. \& {Heiles}, C. 1991, \apj, 382, 204

\bibitem[{{Koo} {et~al.}(1995){Koo}, {Kim}, \& {Seward}}]{1995ApJ...447..211K}
{Koo}, B.-C., {Kim}, K.-T., \& {Seward}, F.~D. 1995, \apj, 447, 211

\bibitem[{{Koo} {et~al.}(2002){Koo}, {Lee}, \& {Seward}}]{2002AJ....123.1629K}
{Koo}, B.-C., {Lee}, J.-J., \& {Seward}, F.~D. 2002, \aj, 123, 1629

\bibitem[{{Koo} {et~al.}(2005){Koo}, {Lee}, {Seward}, \&
  {Moon}}]{2005ApJ...633..946K}
{Koo}, B.-C., {Lee}, J.-J., {Seward}, F.~D., \& {Moon}, D.-S. 2005, \apj, 633,
  946

\bibitem[{{Koo} \& {Moon}(1997{\natexlab{a}})}]{1997ApJ...475..194K}
{Koo}, B.-C. \& {Moon}, D.-S. 1997{\natexlab{a}}, \apj, 475, 194

\bibitem[{{Koo} \& {Moon}(1997{\natexlab{b}})}]{1997ApJ...485..263K}
{Koo}, B.-C. \& {Moon}, D.-S. 1997{\natexlab{b}}, \apj, 485, 263

\bibitem[{{Lopez} {et~al.}(2013){Lopez}, {Pearson}, {Ramirez-Ruiz}, {Castro},
  {Yamaguchi}, {Slane}, \& {Smith}}]{2013ApJ...777..145L}
{Lopez}, L.~A., {Pearson}, S., {Ramirez-Ruiz}, E., {et~al.} 2013, \apj, 777,
  145

\bibitem[{{Mufson} \& {Liszt}(1979)}]{1979ApJ...232..451M}
{Mufson}, S.~L. \& {Liszt}, H.~S. 1979, \apj, 232, 451

\bibitem[{{Nomoto} {et~al.}(1997){Nomoto}, {Hashimoto}, {Tsujimoto},
  {Thielemann}, {Kishimoto}, {Kubo}, \& {Nakasato}}]{1997NuPhA.616...79N}
{Nomoto}, K., {Hashimoto}, M., {Tsujimoto}, T., {et~al.} 1997, Nuclear Physics
  A, 616, 79

\bibitem[{{Seward}(1990)}]{1990ApJS...73..781S}
{Seward}, F.~D. 1990, \apjs, 73, 781

\bibitem[{{Snowden} {et~al.}(2004){Snowden}, {Collier}, \&
  {Kuntz}}]{2004ApJ...610.1182S}
{Snowden}, S.~L., {Collier}, M.~R., \& {Kuntz}, K.~D. 2004, \apj, 610, 1182

\bibitem[{{Snowden} {et~al.}(2008){Snowden}, {Mushotzky}, {Kuntz}, \&
  {Davis}}]{2008A&A...478..615S}
{Snowden}, S.~L., {Mushotzky}, R.~F., {Kuntz}, K.~D., \& {Davis}, D.~S. 2008,
  \aap, 478, 615

\bibitem[{{Str{\"u}der} {et~al.}(2001){Str{\"u}der}, {Briel}, {Dennerl},
  {Hartmann}, {Kendziorra}, {Meidinger}, {Pfeffermann}, {Reppin}, {Aschenbach},
  {Bornemann}, {Br{\"a}uninger}, {Burkert}, {Elender}, {Freyberg}, {Haberl},
  {Hartner}, {Heuschmann}, {Hippmann}, {Kastelic}, {Kemmer}, {Kettenring},
  {Kink}, {Krause}, {M{\"u}ller}, {Oppitz}, {Pietsch}, {Popp}, {Predehl},
  {Read}, {Stephan}, {St{\"o}tter}, {Tr{\"u}mper}, {Holl}, {Kemmer}, {Soltau},
  {St{\"o}tter}, {Weber}, {Weichert}, {von Zanthier}, {Carathanassis}, {Lutz},
  {Richter}, {Solc}, {B{\"o}ttcher}, {Kuster}, {Staubert}, {Abbey}, {Holland},
  {Turner}, {Balasini}, {Bignami}, {La Palombara}, {Villa}, {Buttler},
  {Gianini}, {Lain{\'e}}, {Lumb}, \& {Dhez}}]{2001A&A...365L..18S}
{Str{\"u}der}, L., {Briel}, U., {Dennerl}, K., {et~al.} 2001, \aap, 365, L18

\bibitem[{{Subrahmanyan} \& {Goss}(1995)}]{1995MNRAS.275..755S}
{Subrahmanyan}, R. \& {Goss}, W.~M. 1995, \mnras, 275, 755

\bibitem[{{Turner} {et~al.}(2001){Turner}, {Abbey}, {Arnaud}, {Balasini},
  {Barbera}, {Belsole}, {Bennie}, {Bernard}, {Bignami}, {Boer}, {Briel},
  {Butler}, {Cara}, {Chabaud}, {Cole}, {Collura}, {Conte}, {Cros}, {Denby},
  {Dhez}, {Di Coco}, {Dowson}, {Ferrando}, {Ghizzardi}, {Gianotti}, {Goodall},
  {Gretton}, {Griffiths}, {Hainaut}, {Hochedez}, {Holland}, {Jourdain},
  {Kendziorra}, {Lagostina}, {Laine}, {La Palombara}, {Lortholary}, {Lumb},
  {Marty}, {Molendi}, {Pigot}, {Poindron}, {Pounds}, {Reeves}, {Reppin},
  {Rothenflug}, {Salvetat}, {Sauvageot}, {Schmitt}, {Sembay}, {Short},
  {Spragg}, {Stephen}, {Str{\"u}der}, {Tiengo}, {Trifoglio}, {Tr{\"u}mper},
  {Vercellone}, {Vigroux}, {Villa}, {Ward}, {Whitehead}, \&
  {Zonca}}]{2001A&A...365L..27T}
{Turner}, M. J.~L., {Abbey}, A., {Arnaud}, M., {et~al.} 2001, \aap, 365, L27

\end{thebibliography}

\end{document}